\documentstyle[12pt,epsfig]{article}
\newcommand{\zp}[3]{Z.\ Phys.\ {\bf C#1} (19#2) #3}
\newcommand{\pl}[3]{Phys.\ Lett.\ {\bf B#1} (19#2) #3}
\newcommand{\plold}[3]{Phys.\ Lett.\ {\bf #1B} (19#2) #3}

\newcommand{\prd}[3]{Phys.\ Rev.\ {\bf D#1} (19#2) #3}
\newcommand{\prl}[3]{Phys.\ Rev.\ Lett.\ {\bf #1} (19#2) #3}

\newcommand{\mpl}[3]{Mod.\ Phys.\ Lett.\ {\bf A#1} (19#2) #3}

\newcommand{\md}{\mbox{d}}
\def\simgt{\rlap{\lower 3.5 pt \hbox{$\mathchar \sim$}} \raise 1pt \hbox {$>$}}
\def\simlt{\rlap{\lower 3.5 pt \hbox{$\mathchar \sim$}} \raise 1pt \hbox {$<$}}

\newcommand{\beq}{\begin{equation}}
\newcommand{\eeq}{\end{equation}}
\newcommand{\bea}{\begin{eqnarray}}
\newcommand{\eea}{\end{eqnarray}}

\newcommand{\alps}{\mbox{$\alpha_{\mbox{\scriptsize s}}$}}

\newcommand{\uone}{\mbox{$\underline{1}$}}
\newcommand{\ueight}{\mbox{$\underline{8}$}}

\catcode`\@=11

\def\section{\@startsection{section}{1}{\z@}{3.5ex plus 1ex minus .2ex}
{2.3ex plus .2ex}{\large\bf}}
\def\thesection{\arabic{section}.}

\def\appendix{\setcounter{section}{0}
\def\thesection{Appendix \Alph{section}:}
\def\theequation{\Alph{section}.\arabic{equation}}}
\def\@citex[#1]#2{\if@filesw\immediate\write\@auxout{\string\citation{#2}}\fi
  \def\@citea{}\@cite{\@for\@citeb:=#2\do
    {\@citea\def\@citea{,\penalty\@m}\@ifundefined
       {b@\@citeb}{{\bf ?}\@warning
       {Citation `\@citeb' on page \thepage \space undefined}}%
\hbox{\csname b@\@citeb\endcsname}}}{#1}}
\def\citer{\@ifnextchar [{\@tempswatrue\@citexr}{\@tempswafalse\@citexr[]}}
\def\@citexr[#1]#2{\if@filesw\immediate\write\@auxout{\string\citation{#2}}\fi
  \def\@citea{}\@cite{\@for\@citeb:=#2\do
    {\@citea\def\@citea{--\penalty\@m}\@ifundefined
       {b@\@citeb}{{\bf ?}\@warning
       {Citation `\@citeb' on page \thepage \space undefined}}%
\hbox{\csname b@\@citeb\endcsname}}}{#1}}
\relax
\begin{document}
\thispagestyle{empty}
\begin{flushright}
DESY 96-005\\
January 1996
\end{flushright}
\vskip 2.5cm
\begin{center}
\boldmath
{\Large \bf Color-Octet Contributions to $J/\psi$ Photoproduction}
\unboldmath
\vglue 1.5cm
\begin{sc}
{\large\sc Matteo Cacciari\footnote{e-mail: cacciari@desy.de} 
and Michael Kr\"{a}mer\footnote{e-mail: mkraemer@desy.de}}
\vglue 0.3cm
\end{sc}
{\em Deutsches Elektronen-Synchrotron DESY\\
D-22603 Hamburg, Germany}
\end{center}
\vglue 1.7cm

\begin{abstract} We have calculated the leading color-octet contributions to
  the production of $J\!/\!\psi$ particles in photon-proton
  collisions.  Using the values for the color-octet matrix elements
  extracted from fits to prompt $J\!/\!\psi$ data at the Tevatron,
  we demonstrate that distinctive color-octet signatures should be
  visible in $J\!/\!\psi$ photoproduction. However, these predictions
  appear at variance with recent experimental data obtained at
  \mbox{HERA}, indicating that the phenomenological importance of the
  color-octet contributions is smaller than expected from theoretical
  considerations and suggested by the Tevatron fits.\\ 
  \vskip-5pt PACS numbers: 12.38.Bx,13.60.Le,14.40.Gx 
\end{abstract}

\vfill
\newpage

The production of heavy quarkonium states in high energy collisions
provides an important tool to study the interplay between perturbative
and non-perturbative QCD dynamics. A rigorous framework for treating
quarkonium production and decays has recently been developed in
Ref.\cite{BBL95}. The factorization approach is based on the use of
non-relativistic QCD (NRQCD) to separate the short-distance parts from
the long-distance matrix elements and explicitly takes into account
the complete structure of the quarkonium Fock space.  This formalism
implies that so-called color-octet processes, in which the heavy-quark
antiquark pair is produced at short distances in a color-octet state
and subsequently evolves non-perturbatively into a physical
quarkonium, should contribute to the cross section.  It has recently
been argued in Refs.\cite{TEV1,TEV2} that quarkonium production in
hadronic collisions at the Tevatron can be accounted for by including
color-octet processes and by adjusting the unknown long-distance
color-octet matrix elements to fit the data.

In order to establish the phenomenological significance of the
color-octet mechanism it is necessary to identify color-octet
contributions in different production processes. Color-octet
production of $J\!/\!\psi$ particles in $e^+e^-$ annihilation and $Z$
decays has been studied in Refs.\cite{COEE}.  In this note, we examine
the color-octet contributions to $J\!/\!\psi$ photoproduction, $\gamma
+ P \to J\!/\!\psi + X$, which proceeds predominantly through
photon-gluon fusion. Elastic/diffractive mechanisms \cite{ELASTIC} and
reducible background processes can be eliminated by applying suitable
cuts \cite{JST92}.  According to the formalism developed in
Ref.\cite{BBL95}, the inclusive production cross section can be
expressed as a sum of terms, each of which factors into a
short-distance coefficient and a long-distance matrix element:
\beq\label{eq_fac}
\md\sigma(\gamma+g \to J\!/\!\psi +X) 
= \sum_n \md\hat{\sigma}(\gamma+g \to c\bar{c}\, [n] + X)\, 
         \langle 0|{\cal{O}}^{J/\psi}\,[n]|0\rangle \, .
\eeq
Here, $\md\hat{\sigma}$ denotes the short-distance cross section for
producing an on-shell $c\bar{c}$-pair in a color, spin and
angular-momentum state labelled by $n$.  The NRQCD matrix elements
$\langle 0 | {\cal{O}}^{J\!/\!\psi} \, [n] | 0 \rangle$ give the
probability for a $c\bar{c}$-pair in the state $n$ to form the
$J\!/\!\psi$ particle. The relative importance of the various terms
in (\ref{eq_fac}) can be estimated by using NRQCD velocity scaling
rules \cite{LMNMH92}. For $v\to 0$ ($v$ being the average velocity of
the charm quark in the $J\!/\!\psi$ rest frame) each of the NRQCD
matrix elements scales with a definite power of $v$ and the general
expression (\ref{eq_fac}) can be organized into an expansion in powers
of $v^2$.

\begin{figure}[t]
\begin{center}


\epsfig{file=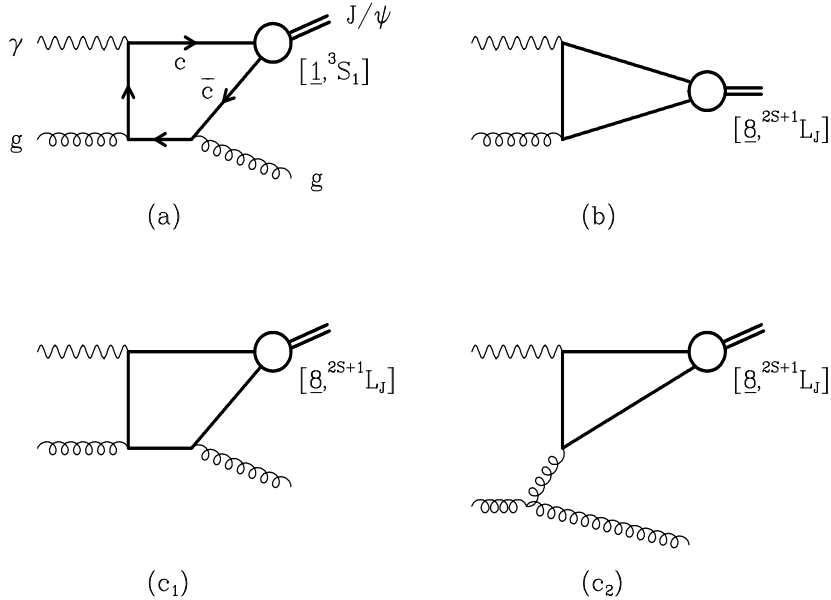,bbllx=10pt,bblly=60pt,bburx=550pt,bbury=770pt,%
        width=10cm,height=12cm,angle=-90}
\end{center}
\caption[dummy]{\label{fig_1} Generic diagrams for $J\!/\!\psi$ 
  photoproduction: (a) leading color-singlet contribution; (b) leading
  color-octet contributions; (c) color-octet contributions to
  inelastic $J\!/\!\psi$ production.}

\end{figure}

At leading order in $v^2$, eq.(\ref{eq_fac}) reduces to the standard
factorization formula of the color-singlet model \cite{CS}. The
short-distance cross section is given by the subprocess
\beq\label{eq_cs}
\gamma + g \to c\bar{c}\, [\uone,{}^3S_{1}] + g
\eeq
shown in Fig.\ref{fig_1}a, with $c\bar{c}$ in a color-singlet state
(denoted by \uone), zero relative velocity, and spin/angular-momentum
quantum numbers $^{2S+1}L_J = {}^3S_1$.  
Higher-order QCD corrections to the short-distance process
(\ref{eq_cs}) were found to increase the color-singlet cross section
by more than 50\%, depending in detail on the choice of parameters
\cite{KZSZ94,MK95}.

Color-octet configurations are produced at leading order in $\alps$
through the $2\to 1$ parton processes
\bea\label{eq_oc0}
\gamma + g &\! \to \!& c\bar{c}\, [\ueight,{}^1S_{0}]\nonumber \\
\gamma + g &\! \to \!& c\bar{c}\, [\ueight,{}^3P_{0,2}]
\eea
shown in Fig.\ref{fig_1}b. The transition of the color-octet $c\bar{c}
\, [\ueight,{}^{2S+1}L_{J}]$ pair into a physical $J\!/\!\psi$ state
through the emission of non-perturbative gluons is described by the
long-distance matrix elements $\langle 0 | {\cal{O}}^{J\!/\!\psi} \,
[\ueight,{}^{2S+1}L_{J}] | 0 \rangle$.  They have to be obtained from
lattice simulations or measured directly in some production process.
According to the velocity scaling rules of NRQCD, the color-octet
matrix elements should be suppressed by a factor of $v^4$ compared to
the leading color-singlet matrix element \cite{footnote1}. Color-octet
contributions to $J\!/\!\psi$ photoproduction can thus become
important only if the corresponding short-distance cross sections are
enhanced as compared to the color-singlet process.  For the partonic
cross section of the leading color-octet contributions we find

\bea\label{eq_oc1}
\hat\sigma(\gamma+g\to c\bar{c}\,[n]\to J\!/\!\psi) =
    \frac{\pi\,\delta(\hat{s}-4m_{c}^{2})}{4m_{c}^{2}}\;     
    \overline{\sum}
    |{\cal{M}}(\gamma+g\to c\bar{c}\,[n])|^2 \,\,  
    \langle 0|{\cal{O}}^{J\!/\!\psi}\,[n]|0\rangle
\eea
with 
\bea\label{eq_ocmsq}
\overline{\sum}|{\cal{M}}(\gamma+g\to c\bar{c}\,[\ueight,{}^{1}S_{0}])|^2  
& \! = \! & \frac{(4\pi)^2 e_c^2 \alpha\alps}{4m_{c}}     
                                                          \nonumber \\ 
\overline{\sum}|{\cal{M}}(\gamma+g\to c\bar{c}\,[\ueight,{}^{3}P_{0}])|^2  
& \! = \! & \frac{3(4\pi)^2 e_c^2 \alpha\alps}{4m_{c}^{3}} 
                                                          \nonumber \\
\overline{\sum}|{\cal{M}}(\gamma+g\to c\bar{c}\,[\ueight,{}^{3}P_{2}])|^2  
& \! = \! & \frac{(4\pi)^2 e_c^2 \alpha\alps}{5m_{c}^{3}} \, . 
\eea
The partonic energy squared is denoted by $\hat{s}$, and $e_c$ is the
magnitude of the charm quark charge in units of $e=\sqrt{4\pi\alpha}$.
We have checked that our results are consistent with those presented
in Refs.\cite{TEV2,FM95}.

In the analyses Refs.\cite{TEV1,TEV2}, color-octet matrix elements
have been fitted to prompt $J\!/\!\psi$ data from \mbox{CDF}
\cite{CDF} and found to be consistent with the NRQCD velocity scaling
rules.  The fit-values have, however, large theoretical errors due to
the uncertainty in the heavy quark mass and the QCD coupling, unknown
higher-order corrections, and higher twist effects \cite{HT}, and
should therefore mainly be regarded as order-of-magnitude estimates.
For our numerical analysis we choose the color-octet matrix elements
as listed in Table~\ref{table1}, consistent with the velocity scaling
rules and the results obtained in Refs.\cite{TEV1,TEV2}.  The
color-singlet matrix element has been calculated from the $J\!/\!\psi$
wave function at the origin as obtained in the QCD-motivated potential
model of Ref.\cite{BT81} and tabulated in Ref.\cite{EQ95}. At leading
order in $v^2$, the $P$-wave matrix elements are related by
heavy-quark spin symmetry, $\langle 0 | {\cal{O}}^{J\!/\!\psi}
\,[\ueight,{}^{3}P_{J}] | 0 \rangle \approx \mbox{$(2J+1)$} \, \langle
0 | {\cal{O}}^{J\!/\!\psi} \, [\ueight,{}^{3}P_{0}] | 0 \rangle$.

\begin{table}[htbp]

\begin{center}
\begin{tabular}{|c|c|c|} \hline 
\rule[-2.5mm]{0mm}{7mm}
$\langle 0|{\cal{O}}^{J/\psi}\,[n]|0\rangle$ & numerical value & scaling\\
\hline 
\rule[-2.5mm]{0mm}{7mm}
$\langle 0|{\cal{O}}^{J/\psi}\,[\uone,{}^{3}S_{1}]|0\rangle
\hphantom{/m_c^2}$ 
& $1.16$ GeV$^3$ & $m_c^3 v^3$ \\ \rule[-2.5mm]{0mm}{7mm}
$\langle 0|{\cal{O}}^{J/\psi}\,[\ueight,{}^{1}S_{0}]|0\rangle
\hphantom{/m_c^2}$ 
& $10^{-2}$ GeV$^3$ & $m_c^3 v^7$ \\ \rule[-2.5mm]{0mm}{7mm}
$\langle 0|{\cal{O}}^{J/\psi}\,[\ueight,{}^{3}P_{0}]|0\rangle / m_c^2$ 
& $10^{-2}$ GeV$^3$ & $m_c^3 v^7$ \\ 
\hline
\end{tabular}
\caption[dummy]{\label{table1} Values of the NRQCD matrix elements used 
  in the numerical analysis, with the velocity and mass scaling.}

\end{center}

\vspace*{-3mm}

\end{table}

\begin{figure}[t]

\begin{center}
\epsfig{file=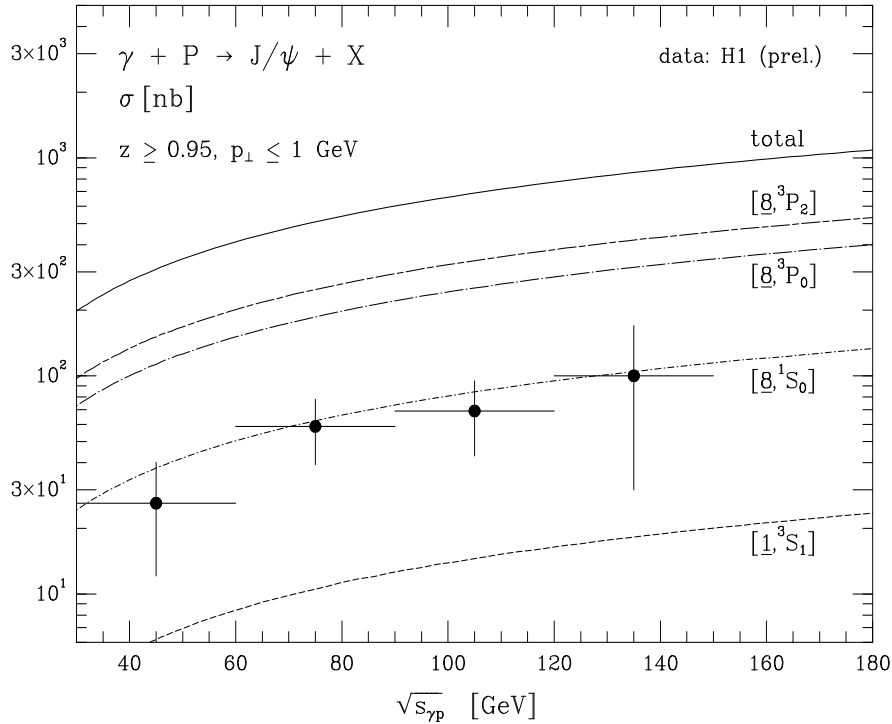,bbllx=10pt,bblly=60pt,bburx=550pt,bbury=770pt,%
        width=10cm,height=12cm,angle=-90}
\end{center}

\vspace*{2mm}

\caption[dummy]{\label{fig_2} Total cross section for $J\!/\!\psi$ 
  photoproduction in the region $z\ge 0.95$ and $p_\perp \le 1$~GeV as
  a function of the photon-proton energy. Experimental data from
  \cite{H1}.}

\end{figure}

Due to kinematical constraints, the leading color-octet terms will
only contribute to the upper endpoint of the $J\!/\!\psi$ energy
spectrum, $z\approx 1$ and $p_\perp\approx 0$. The $J\!/\!\psi$ energy
variable is defined by $z = {p\cdot k_\psi}\, / \, {p\cdot k_\gamma}$,
with $p, k_{\psi,\gamma}$ being the momenta of the proton and
$J\!/\!\psi$, $\gamma$ particles, respectively, and $p_\perp$ is the
$J\!/\!\psi$ transverse momentum. The leading color-octet and
color-singlet contributions to the $J\!/\!\psi$ photoproduction cross
section in the region $z \ge 0.95$ and $p_\perp \le 1$~GeV are shown
in Fig.\ref{fig_2}. The predictions are compared to experimental data
obtained at \mbox{HERA} by the \mbox{H1} collaboration \cite{H1}.
We have used $m_c = 1.48$~\cite{BT81}, $\alps(2m_c) = 0.3$, and the
values for the NRQCD matrix elements listed in Table \ref{table1}.
The parton cross section has been convoluted with the leading-order
GRV parametrization of the gluon density in the proton \cite{GRV95},
evaluated at a scale $Q=2m_c$. Relativistic corrections to the
color-singlet channel \cite{REL} enhance the large $z$ region, but
cannot change the order-of-magnitude suppression of the color-singlet
process significantly. The large cross section predicted by the
color-octet mechanism appears to be in conflict with the experimental
data.  This indicates that the value of the color-octet matrix
elements $\langle 0 | {\cal{O}}^{J\!/\!\psi} \, [\ueight,{}^{1}S_{0}]
| 0 \rangle$ and $ \langle 0 | {\cal{O}}^{J\!/\!\psi} \,
[\ueight,{}^{3}P_{J}] | 0 \rangle$ is more than one order of magnitude
smaller than expected from the velocity scaling rules and suggested by
the Tevatron fits. It is, however, difficult to put strong upper
limits for the octet terms from a measurement of the total cross
section since the overall normalization of the theoretical prediction
depends strongly on the choice for the charm quark mass and the QCD
coupling.  Moreover, diffractive production mechanisms which cannot be
calculated within perturbative QCD might contaminate the region
$z\approx 1$ and make it difficult to extract precise information on
the color-octet contributions.

Diffractive processes can be eliminated by restricting the analysis to
the inelastic domain $z \le 0.9$ and $p_\perp \ge 1$~GeV.  Color-octet
configurations which contribute to inelastic $J\!/\!\psi$
photoproduction are produced through the subprocesses
\bea\label{eq_oc2}
\gamma + g &\! \to \!& c\bar{c}\, [\ueight,{}^1S_{0}] + g \nonumber \\
\gamma + g &\! \to \!& c\bar{c}\, [\ueight,{}^3S_{1}] + g \nonumber \\
\gamma + g &\! \to \!& c\bar{c}\, [\ueight,{}^3P_{0,1,2}] + g 
\eea
as shown in Fig.\ref{fig_1}c. Light-quark initiated processes are
strongly suppressed at \mbox{HERA} energies and can safely be
neglected in a first step.  The parton cross sections (\ref{eq_oc2})
have been evaluated using the algebraic computer program FORM
\cite{form}. Details of the calculation and analytic results will be
presented in a forthcoming publication \cite{CK96}.  Here we merely
note that the cross sections $\gamma + g \to c\bar{c}\, [\ueight,
{}^1S_{0}], [\ueight, {}^3P_{0,2}] + g$ are divergent for $z\to 1$ and
$p_\perp \to 0$, due to the $g\to gg$ collinear splitting,
Fig.\ref{fig_1}c$_2$. This singularity can be absorbed into the
renormalization of the parton densities via mass factorization
\cite{CK96}. In the inelastic region $z \le 0.9$ and $p_\perp \ge
1$~GeV the cross sections are well-behaved however.

\begin{figure}[t]

\begin{center}
\epsfig{file=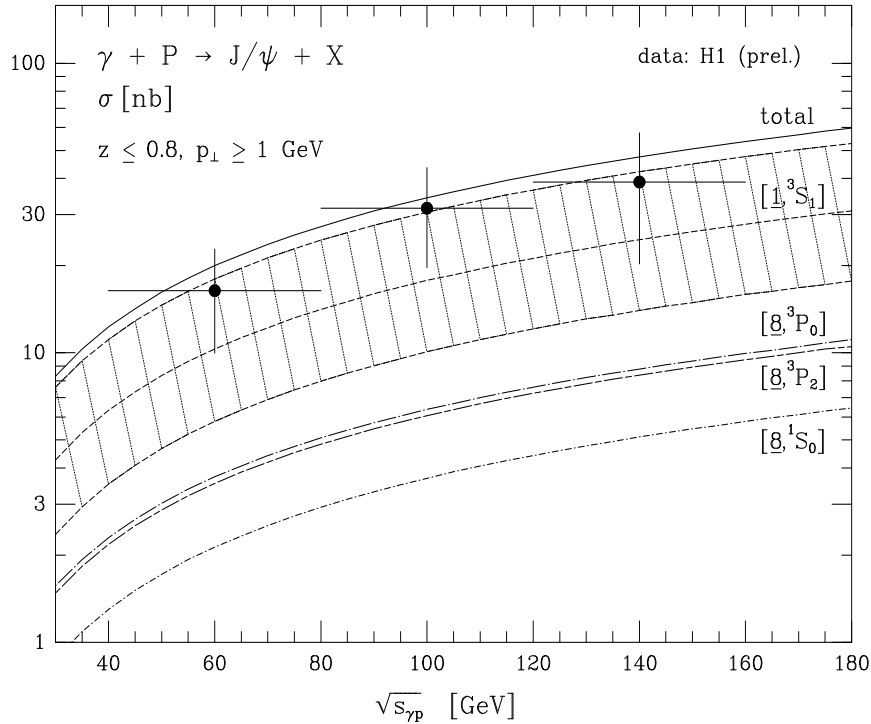,bbllx=10pt,bblly=60pt,bburx=550pt,bbury=770pt,%
        width=10cm,height=12cm,angle=-90}
\end{center}

\vspace*{2mm}

\caption[dummy]{\label{fig_3} Total cross section for $J\!/\!\psi$ 
  photoproduction in the region $z\le 0.8$ and $p_\perp \ge 1$~GeV as
  a function of the photon-proton energy. Experimental data from
  \cite{H1}.}

\end{figure}

In Fig.\ref{fig_3} we show the color-octet and (leading-order)
color-singlet contributions to inelastic $J\!/\!\psi$ photoproduction
compared to experimental data from \mbox{H1} \cite{H1} obtained in the
kinematical region $z\le 0.8$ and $p_\perp \ge 1$~GeV. For the central
predictions we have used $m_c=1.48$~GeV and $\alps=0.3$, while the
hatched error band indicates how much the color-singlet cross section
is altered if $m_c$ and $\alps$ vary in the range $1.4$~GeV $\le m_c
\le$ $1.55$~GeV and $0.25 \le \alps \le 0.35$.  Adopting the NRQCD
matrix elements listed in Table \ref{table1} and $\langle 0 |
{\cal{O}}^{J\!/\!\psi} \, [\ueight,{}^{3}S_{1}] | 0 \rangle =
10^{-2}$~GeV$^3$ \cite{TEV1,TEV2} we find that color-octet and
color-singlet contributions to the inelastic cross section are
predicted to be of comparable size.  The short-distance factors of the
$[\ueight,{}^{1}S_{0}]$ and $[\ueight,{}^{3}P_{0,2}]$ channels are
strongly enhanced as compared to the color-singlet term and partly
compensate the ${\cal{O}}(10^{-2})$ suppression of the corresponding
non-perturbative matrix elements.  In contrast, the contributions from
the $[\ueight,{}^{3}S_{1}]$ and $[\ueight,{}^{3}P_{1}]$ states are
suppressed by more than one order of magnitude.  Since color-octet and
color-singlet processes contribute at the same order in $\alps$, the
large size of the $[\ueight,{}^{1}S_{0}]$ and
$[\ueight,{}^{3}P_{0,2}]$ cross sections could not have been
anticipated from naive
power counting and demonstrates the crucial dynamical role played by
the bound state quantum numbers \cite{BR83}.  Inclusion of the
color-octet processes increases the inelastic cross section by about a
factor of two, consistent with experimental data.  This does, however,
not proof the significance of the color-octet terms: as can be seen
from Fig.\ref{fig_3}, the experimental data can be accounted for by
the color-singlet process alone, once the theoretical uncertainties
due to variation of the charm quark mass and the strong coupling are
properly taken into account. This has also been demonstrated in the
next-to-leading order analysis of Ref.\cite{MK95}.  No strong
conclusions on the size of the color-octet terms can thus be deduced
from the study of the total inelastic cross section. The same
statement holds true for the transverse momentum spectrum, since, at
small and moderate $p_\perp$, both color-singlet and color-octet
contributions are almost identical in shape.  At large transverse
momenta, $p_\perp \; \simgt \; 10$~GeV, charm quark fragmentation
dominates over the photon-gluon fusion process \cite{SA94,GRS95}.  In
contrast to what was found at the Tevatron \cite{PT_TEV}, gluon
fragmentation into color-octet states is suppressed over the whole
range of $p_\perp$ \cite{GRS95}.

\begin{figure}[t]

\begin{center}
\epsfig{file=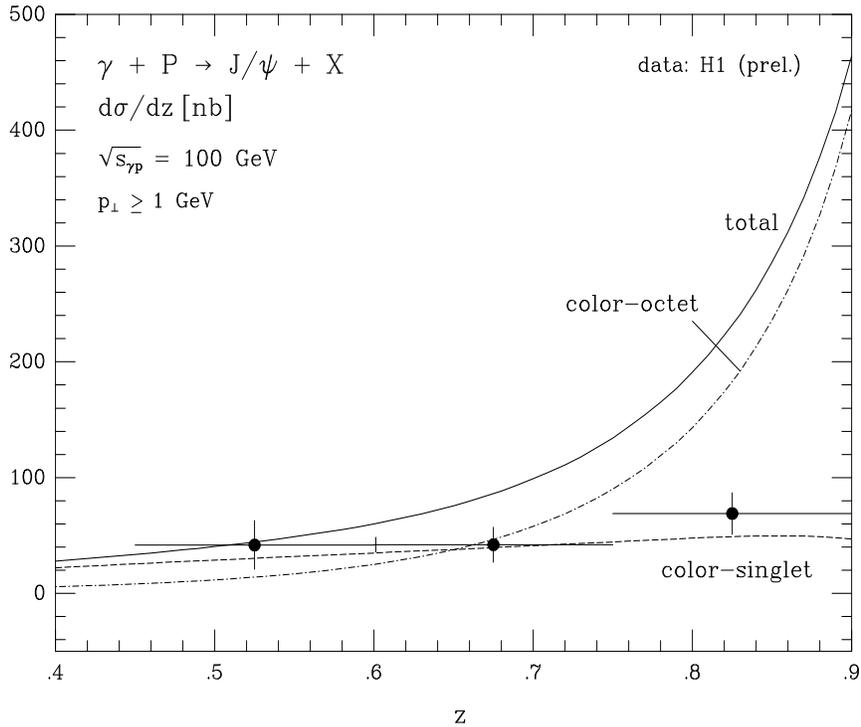,bbllx=10pt,bblly=60pt,bburx=550pt,bbury=770pt,%
        width=10cm,height=12cm,angle=-90}
\end{center}

\vspace*{2mm}

\caption[dummy]{\label{fig_4} The $J\!/\!\psi$ energy distribution 
  $\md\sigma/\md{}z$ at the photon-proton centre of mass energy
  $\sqrt{s\hphantom{tk}}\!\!\!\!\!  _{\gamma p}\,\, = 100$~GeV
  integrated in the range $p_\perp \ge 1$~GeV.  Experimental data from
  \cite{H1Z}.}

\end{figure}

A distinctive signal for color-octet processes should, however, be
visible in the $J\!/\!\psi$ energy distribution $\md\sigma/\md{}z$
shown in Fig.\ref{fig_4}.  We have plotted color-singlet and
color-octet contributions at a typical \mbox{HERA} energy of
$\sqrt{s\hphantom{tk}} \!\!\!\!\!  _{\gamma p}\,\, = 100$~GeV in the
restricted range $p_\perp \ge 1$~GeV, compared to recent experimental
data from \mbox{H1} \cite{H1Z}.  Since the shape of the distribution
is insensitive to higher-order QCD corrections or to the uncertainty
induced by the choice for $m_c$ and $\alps$, the analysis of the
$J\!/\!\psi$ energy spectrum $\md\sigma/\md{}z$ provides a clean test
for the underlying production mechanism. The comparison with the
experimental data, Fig.\ref{fig_4}, shows that the $J\!/\!\psi$ energy
spectrum is adequately accounted for by the color-singlet
contribution.  The shape predicted by the color-octet contributions is
instead in conflict with the experimental data. 

In conclusion, we have investigated color-octet contributions to the
production of $J\!/\!\psi$ particles in photon-proton collisions.
Color-octet processes would strongly affect the upper endpoint region
of the $J\!/\!\psi$ energy spectrum and the shape of the $J\!/\!\psi$
energy distribution. However, these predictions appear at variance
with recent experimental data obtained at \mbox{HERA}. We conclude
that the values of the color-octet matrix elements
$\langle 0 | {\cal{O}}^{J\!/\!\psi} \, [\ueight,{}^{1}S_{0}] | 0
\rangle$ and $ \langle 0 | {\cal{O}}^{J\!/\!\psi} \,
[\ueight,{}^{3}P_{J}] | 0 \rangle$ seem to be more than one order of
magnitude smaller than expected from the velocity scaling rules and
suggested by the Tevatron fits.  With higher statistics data it will
be possible to extract more detailed information on the color-octet
matrix elements, in particular from the analysis of the $J\!/\!\psi$
energy distribution in the inelastic region.

\vspace*{2mm}

\noindent
{\bf Acknowledgements}

\noindent 
It is a pleasure to thank Mario Greco and Peter M.\ 
Zerwas for suggestions and comments. Special thanks go to the
\mbox{H1} collaboration, in particular to Beate Naroska, Stephan
Schiek and Guido Schmidt, for providing experimental data prior to
publication.  We have benefitted from conversations with Riccardo
Brugnera, Peppe Iacobucci, Wolfgang Kilian, Luca Stanco, Mikko
V\"anttinen and Paolo Vitulo.



\begin{thebibliography}{999}

\bibitem{BBL95} G.T.~Bodwin, E.~Braaten, and G.P.~Lepage, \prd{51}{95}{1125}.

\bibitem{TEV1} E.~Braaten and S.~Fleming, \prl{74}{95}{3327}; 
               M.~Cacciari, M.~Greco, M.L.~Mangano, and A.~Petrelli, 
               \pl{356}{95}{560}.

\bibitem{TEV2} P.~Cho and A.K.~Leibovich, CALT-68-1988 and CALT-68-2026. 

\bibitem{COEE} E.~Braaten and Y.-Q.~Chen, NUHEP-TH-95-9;
               K.~Cheung, W.-Y.~Keung, and T.C.~Yuan, FERMILAB-PUB-95/300-T;
               P.~Cho, CALT-68-2020.

\bibitem{ELASTIC} A.~Donnachie and P.V.~Landshoff, \pl{348}{95}{213}; 
         M.G.~Ryskin, R.G.~Roberts, A.D.~Martin, and E.M.~Levin, DTP-95-96;
         T.~Ahmed et al.\ [H1 Collab.], \pl{338}{94}{507};
         M.~Derrick et al.\ [ZEUS Collab.], \pl{350}{95}{120}.

\bibitem{JST92} H.~Jung, G.A.~Schuler, and J.~Terr\'{o}n,
                Int.~J.~Mod.~Phys.\ {\bf A7} (1992) 7955. 

\bibitem{LMNMH92} G.P.~Lepage, L.~Magnea, C.~Nakhleh, U.~Magnea, and  
                  K.~Hornbostel, \prd{46}{92}{4052}.

\bibitem{CS} E.L.~Berger and D.~Jones, \prd{23}{81}{1521}; 
             R.~Baier and R.~R\"{u}ckl, \plold{102}{81}{364}.

\bibitem{KZSZ94} M.~Kr\"{a}mer, J.~Zunft, J.~Steegborn, and P.M.~Zerwas, 
                 \pl{348}{95}{657}.

\bibitem{MK95} M.~Kr\"{a}mer, DESY-95-155, Nucl.~Phys.~B.\ in press.

\bibitem{footnote1}{In the case of $P$-wave quarkonia, color-singlet 
      and color-octet matrix elements contribute at the same order in $v$, 
      see: G.T.~Bodwin, E.~Braaten, and G.P.~Lepage, \prd{46}{92}{R1914}.
      Photoproduction of $P$-wave states has been studied in: 
      J.P.~Ma, UM-P-95-96.}

\bibitem{FM95} S.~Fleming and I.~Maksymyk, MADPH-95-922.

\bibitem{CDF} F.~Abe et al. [CDF Collab.], \prl{69}{92}{3704};
              A.~Sansoni, [CDF Collab.], FERMILAB-CONF-95/263-E.

\bibitem{HT} M.~V\"{a}nttinen, P.~Hoyer, S.J.~Brodsky, and W.K.~Tang, 
             \prd{51}{95}{3332}; M.\ V\"{a}nttinen and W.K.\ Tang, 
             SLAC-PUB-95-6931.

\bibitem{BT81} W.~Buchm\"{u}ller and S.-H.H.~Tye, \prd{24}{81}{132}.

\bibitem{EQ95} E.J.~Eichten and C.~Quigg, \prd{52}{95}{1726}. 

\bibitem{H1}  S.~Aid et al.\ [H1 Collab.], contribution 
              to the Int.\ Europhysics Conf.\ on High Energy Physics, 
              Brussels, 1995, and publication in preparation. 

\bibitem{GRV95} M.~Gl\"{u}ck, E.~Reya, and A.~Vogt, \zp{67}{95}{433}.

\bibitem{REL} W.Y.~Keung and I.J.~Muzinich, \prd{27}{83}{1518};
              H.~Jung, D.~Kr\"{u}cker, C.~Greub, and D.~Wyler,  
              \zp{60}{93}{721}, H.~Khan and P.~Hoodbhoy, hep-ph 9511360. 

\bibitem{form} FORM~2.0 by J.A.M.~Vermaseren, CAN, Amsterdam, 1991.


\bibitem{CK96} M.~Cacciari and M.~Kr\"amer, in preparation.  

\bibitem{BR83} R.~Baier and R.~R\"{u}ckl, \zp{19}{83}{251}.

\bibitem{SA94} V.A.~Saleev, \mpl{9}{94}{1083}.

\bibitem{GRS95} R.~Godbole, D.P.~Roy, and K.~Sridhar, TIFR/TH/95-57.

\bibitem{PT_TEV} M.~Cacciari and M.~Greco, \prl{73}{94}{1586};
                 E.~Braaten, M.A.~Doncheski, S.~Fleming, and M.L.~Mangano, 
                 \pl{333}{94}{548}; D.P.~Roy and K.~Sridhar, 
                 \pl{339}{94}{141}.

\bibitem{H1Z} H1 Collaboration, private communication (data analysis as 
              described in Ref.\cite{H1}).

\end{thebibliography}
\end{document}